\documentclass{pasa}%
\usepackage{longtable}

\title[The unusual photometric variability of the PMS star GM Cep]{The unusual photometric variability of the PMS star GM Cep}
\author[Semkov et al.]{E. H. Semkov$^1$, S. I. Ibryamov$^1$, S. P. Peneva$^1$, T. R. Milanov$^2$, K. A. Stoyanov$^1$, I. K. Stateva$^1$, D. P. Kjurkchieva$^2$, D. P. Dimitrov$^1$ \and V. S. Radeva$^2$\\
\affil{$^1$Institute of Astronomy and National Astronomical Observatory, Bulgarian Academy of Sciences,
              72 Tsarigradsko Shose blvd., BG-1784 Sofia, Bulgaria}%
\affil{$^2$Department of Physics, Shumen University, 9700 Shumen, Bulgaria}}%
\jid{PASA}
\doi{10.1017/pas.\the\year.xxx}
\jyear{\the\year}

\begin{document}

\begin{abstract}
Results from $UBVRI$ photometric observations of the pre-main sequence star GM Cep obtained in the period April 2011 - August 2014 are reported in the paper.
Presented data are a continuation of our photometric monitoring of the star started in 2008.
GM Cep is located in the field of the young open cluster Trumpler 37 and over the past years it has been an object of intense photometric and spectral studies.
The star shows a strong photometric variability interpreted as a possible outburst from EXor type in previous studies.
Our photometric data for a period of over six years show a large amplitude variability ($\Delta V$ $\sim$ 2.3 mag) and several deep minimums in brightness are observed.
The analysis of the collected multicolor photometric data shows the typical of UX Ori variables a color reversal during the minimums in brightness.
The observed decreases in brightness have a different shape, and evidences of periodicity are not detected.
At the same time, high amplitude rapid variations in brightness typical for the classical T Tauri stars also present on the light curve of GM Cep.
The spectrum of GM Cep shows the typical of classical T Tauri stars wide H$\alpha$ emission line and absorption lines of some metals.
We calculate the outer radius of the H$\alpha$ emitting region as 10.4$\pm$0.5 R$_{\odot}$ and the accretion rate as 1.8$\times$10$^{â-7}$~M$_{\odot}$~yr$^{-1}$.
\end{abstract}
\maketitle%
\section{Introduction}

Photometric variability is a fundamental characteristic of the pre-main sequence (PMS) stars, which manifests as transient increases in brightness (outbursts), temporary drops in brightness (eclipses), irregular or regular variations for a short or long time scales.
Both types of PMS stars the widespread low-mass ($\it M$ $\leq$ $2M_{\odot}$) T Tauri Stars (TTSs) and the more massive Herbig Ae/Be (HAEBE) stars  indicate photometric variability with various amplitudes and periods \cite{her94,her07}.
The TTSs can be separated into two subclasses: Classical T Tauri (CTT) stars surrounded by a massive accretion disk and Weak line T Tauri (WTT) stars without indications of disk accretion \cite{ber}.
According to Herbst et al. \shortcite{her07} the large amplitude variability of CTT stars is caused by magnetically channeled accretion from the circumstellar disk onto the stellar surface.

Some PMS stars show variability in brightness with very large amplitudes, dominated by fading or bursting behavior.
The large amplitude outbursts can be grouped into two main types, named after their respective prototypes: FU Orionis (FUor) and EX Lupi (EXor) \cite{ra10}.  
Both types of eruptive stars seems to be related to young stellar objects with massive circumstellar disks, and their outbursts are commonly attributed to a sizable increase in the disc accretion rate onto the stellar surface \cite{har96}.
During the quiescence state FUors and EXors are normally accreting TTSs, but due to thermal or gravitational instability in the circumstellar disk accretion rate enhanced by a few orders of magnitude up to $\sim$10$^{-4}$$M_{\odot}$$/$yr.

A significant part of HAEBE stars and early type CTT stars show strong photometric variability with sudden quasi-Algol drops in brightness and amplitudes up to 2.5 mag. ($V$) \cite{nata97,van98}. 
During the deep minimums of brightness, an increase in polarization and specific color variability (called ``blueing effect") are observed. 
The prototype of this group of PMS objects with intermediate mass named UXors is UX Orionis. 
The widely accepted explanation of its variability is a variable extinction from dust clumps or filaments passing through the line of sight to the star \cite{dul03,grin91}.
Normally the star becomes redder when its light is covered by dust, but when the obscuration rises sufficiently, the part of the scattered light in the total observed light become considerable and the star color gets bluer.

The PMS star GM Cep lie in the field of the young open cluster Trumpler 37 ($\sim$4 Myr old) at a distance of 870 pc \cite{con02} and most likely is a member of the cluster \cite{mar87,sic05}.
The early long-term photographic observations of the star performed by Suyarkova \shortcite{su75} and Kun \shortcite{kun86} indicate for a large amplitude photometric variability (the observed amplitudes are $\Delta$$m_{pg}$=2.2 mag and $\Delta$V=2.15 mag respectively).
A multicolor photometric study based on optical, infrared and millimeter observations of GM Cep was reported by Sicilia-Aguilar et al. \shortcite{sic08}.
The authors found the star much brighter in 2006 than in 1990 and conclude that the most probable explanation for the brightness increase is an EXor type outburst.

According to Sicilia-Aguilar et al. \shortcite{sic08} GM Cep is a PMS star with solar mass ($\it M$ $\sim$ $2.1 M_\odot$) from G7V-K0V spectral type and with radius between 3 and 6 $R_\odot$.
The observed strong IR excesses has been explained by the presence of a very luminous and massive circumstellar disk.  
The H$\alpha$ emission line in the spectrum of GM Cep has a strong P Cyg profile and the equivalent width of the line vary significantly from 6$\AA$ to 19$\AA$ \cite{sic08}. 
A variable accretion rate (up to $\sim$ 10$^{-6}$ $M_\odot$/year) are also detected in the study of Sicilia-Aguilar et al. \shortcite{sic08}.

A long-term photometric study of GM Cep for several decades period was performed by Xiao et al. \shortcite{xiao}.
The photographic plate archives from Harvard College Observatory and from Sonneberg Observatory are used to construct the long-term $B$ and $V$ light  curves of the star.
The results suggest that GM Cep do not show fast rises in brightness typical of EXor variables and the light curves seem to be dominated by dips superposed on the quiescence state.
Evidences for periodicity of observed dips in brightness were not found in the study of Xiao et al. \shortcite{xiao}.

In our first paper \cite{sem12} the results from $BVRI$ optical photometric observations of the star collected in the period June 2008 - February 2011 are reported.
During out photometric monitoring two deep minimums in brightness are observed. 
The collected multicolor photometric data shows the typical of UXor variables a color reversal during the minimums in brightness. 
Chen et al. \shortcite{chen} reported results from intensive $BVR$ photometric monitoring of GM Cep during the period 2009-2011.
They confirm the UXor nature of variability and suggest an early stage of planetesimal formation in the star environment.
Chen \& Hu \shortcite{chenhu} suggest a periodicity of about 300 days at the observed deep declines in brightness.

Recent $BVRI$ CCD photometric observations of GM Cep collected in the period April 2011 - August 2014 are reported in the present paper.
The multicolor observations give us the opportunity to clarify the mechanism of the brightness variations.

\section{Observations}

Our photometric CCD data were obtained in two observatories with four telescopes: the 2-m Ritchey-Chr\'{e}tien-Coud\'{e} (2-m), the 50/70-cm Schmidt (Sch) and the 60-cm Cassegrain (60-cm) telescopes of the National Astronomical Observatory Rozhen (Bulgaria) and the 1.3-m Ritchey-Cr\'{e}tien (1.3-m) telescope of the Skinakas Observatory\footnote{Skinakas Observatory is a collaborative project of the University of Crete, the Foundation for Research and Technology - Hellas, and the Max-Planck-Institut f\"{u}r Extraterrestrische Physik.} of the Institute of Astronomy, University of Crete (Greece).
The technical parameters and chip specifications for the cameras used with the 2-m RCC, the 1.3-m RC and the 50/70-cm Schmidt telescopes are summarized in Semkov \& Peneva \shortcite{sem12}.
Observations with the 60-cm Cassegrain telescope were performed with FLI PL09000 CCD camera ($3056\times3056$ pixels, 12 $\mu$m pixel size, 16.8$\times$16.8 arc. min. field, 8.5 $e^-$rms RON) 
As references, we used the comparison sequence of fifteen stars in the field around GM Cep published in Semkov \& Peneva \shortcite{sem12}.

All frames were taken through a standard Johnson-Cousins set of filters.
Twilight flat fields in each filter were obtained each clear evening.
All frames obtained with the ANDOR and Vers Array cameras are bias subtracted and flat fielded.
CCD frames obtained with the FLI PL16803 and FLI PL09000 cameras are dark subtracted and flat fielded.
Aperture photometry was performed using DAOPHOT routines.
All the data were analyzed using the same aperture, which was chosen as 6 arc sec in radius, while the background annulus was from 10 to 15 arc sec.

A medium-resolution spectrum of GM Cep was obtained on 2008 June 27 with the 1.3-m RC telescope in Skinakas Observatory.
The focal reducer, ISA 608 spectral CCD camera ($2000\times800$ pixels, 15$\times$15$\mu$m pixel size), 1300 lines/mm grating and 160$\mu$m slit were used.
The combination of used CCD camera, slit and grating yield a resolving power $\lambda/\Delta\lambda$ $\sim$ 1300 at H$\alpha$ line.
The exposure of GM Cep were followed immediately by an exposure of an FeHeNeAr comparison lamp.

\onecolumn
\begin{longtable}{llllllllllll}
\caption{Photometric $IRVB$ observations of GM Cep during the period April 2011 - August 2014}\\
\hline\hline
J.D. (24...) & I & R & V & B & Tel & J.D. (24...) & I & R & V & B & Tel\\ 
\noalign{\smallskip}  
\hline
\endfirsthead
\caption{continued.}\\
\hline\hline
\noalign{\smallskip}  
J.D. (24...) & I & R & V & B & Tel & J.D. (24...) & I & R & V & B & Tel\\ 
\noalign{\smallskip}  
\hline
\endhead
\hline\hline
\endfoot
55656.458 &	11.61 & 12.53 &	13.38 & 14.70 & Sch & 55896.222 & 12.62 & 13.79 & 14.70 & 16.00 & Sch\\
55659.492	& 11.73 & 12.62 &	13.47 & 14.87 & 2-m & 55925.200 & 12.74 & 13.90 & 14.87 & 16.14 & Sch\\
55683.557 &	11.83 & 12.76 &	13.68 & 15.10 & 2-m & 55928.207 & 12.41 & 13.57 & 14.54 & 15.94 & Sch\\
55703.359	& 11.67 & 12.65 &	13.48 & 14.78 & Sch & 55957.187 & 12.06 & 13.01 & 13.95 & - & 2-m\\
55704.370	& 11.62 & 12.56 &	13.43 & 14.74 & Sch & 55958.211 & 12.01 & 12.95 & 13.88 & 15.33 & 2-m\\
55705.376	& 11.57 & 12.50 &	13.34 & 14.66 & Sch & 56003.528 & 12.19 & 13.31 & 14.29 & 15.72 & Sch\\
55706.362 &	11.59 & 12.52 &	13.36 & 14.67 & Sch & 56015.536 & 12.22 & 13.26 & 14.28 & 15.74 & 2-m\\
55707.358 &	11.62 & 12.55 &	13.42 & 14.74 & Sch & 56030.460 & 12.12 & 13.19 & 14.13 & 15.50 & Sch\\
55721.357 & 11.71 &	12.64 & 13.54 &	14.95 & 2-m & 56060.390 & 12.19 & 13.34 & 14.31 & 15.68 & Sch\\
55722.396 & 11.62 &	12.56 & 13.45 &	14.81 & Sch & 56068.375 & 12.11 & 13.20 & 14.18 & 15.63 & Sch\\
55734.452 & 11.78 &	12.77 & 13.69 &	15.07 & Sch & 56091.418 & 12.01 & 13.06 & 14.00 & 15.35 & Sch\\
55735.410 & 11.82 &	12.83 & 13.75 &	15.12 & Sch & 56092.406 & 11.97 & 13.00 & 13.93 & 15.30 & Sch\\
55736.407 & 11.98 &	13.06 & 14.00 &	15.35 & Sch & 56094.469 & 12.21 & 13.21 & 14.18 & 15.60 & 2-m\\
55737.425 & 12.02 &	13.10 & 14.05 &	15.39 & Sch & 56096.423 & 11.96 & 12.99 & 13.94 & 15.31 & Sch\\
55739.553 & 11.85 &	12.81 & 13.71 & -     & 60-cm & 56120.397 & 11.84 & 12.78 & 13.70 & 15.06 & Sch\\
55770.389 &	11.84 &	12.86 &	13.78 &	15.14 & Sch & 56121.291 & 11.89 & 12.87 & 13.76 & 15.14 & Sch\\
55785.307 & -     & 12.47 & -     & -     & Sch & 56122.352 & 11.93 & 12.91 & 13.81 & 15.18 & Sch\\
55786.268 & -     & 12.47 & 13.26 & 14.54 & Sch & 56123.416 & 11.98 & 12.95 & 13.86 & 15.23 & Sch\\
55787.286 & -     & 12.37 & 13.19 & 14.45 & Sch & 56137.318 & 11.74 & 12.67 & 13.51 & 14.79 & Sch\\
55788.314 & -     & 12.40 & 13.22 & 14.47 & Sch & 56139.292 & 11.80 & 12.75 & 13.63 & 14.97 & 1.3-m\\
55789.321 & -     & 12.44 & 13.28 & 14.55 & Sch & 56139.305 & 11.81 & 12.79 & 13.62 & 14.92 & Sch\\
55790.250 &	11.52 & 12.38 &	13.21 &	14.53 & 1.3-m & 56141.385 & 11.72 & 12.64 & 13.50 & 14.86 & 1.3-m\\
55790.261 & -     & 12.41 & 13.22 & 14.46 & Sch & 56142.256 & 11.73 & 12.64 & 13.51 & 14.85 & 1.3-m\\
55791.277 &	11.48 &	12.34 &	13.17 &	14.47 & 1.3-m & 56145.555 & 11.62 & 12.54 & 13.34 & 14.49 & 60-cm\\
55791.292 & 11.48 & 12.37 & 13.17 & 14.44 & Sch & 56157.592 & 11.76 & 12.68 & 13.54 & 14.88 & 1.3-m\\
55792.244 &	11.53 & 12.38 &	13.22 &	14.54 & 1.3-m & 56159.371 & 11.67 & 12.56 & 13.42 & 14.76 & Sch\\
55792.279 & 11.52 & 12.43 & 13.23 & 14.50 & Sch & 56160.352 & 11.58 & 12.46 & 13.29 & 14.60 & Sch\\
55797.343 &	11.48 & 12.35 &	13.16 &	14.45 & Sch & 56161.374 & 11.59 & 12.50 & 13.31 & 14.63 & Sch\\
55798.328 &	11.47 & 12.35 & 13.16 &	14.45 & Sch & 56162.357 & 11.65 & 12.55 & 13.38 & 14.69 & Sch\\
55799.342 &	11.51 &	12.41 &	13.24 &	14.52 & Sch & 56166.267 & 11.66 & 12.56 & 13.40 & 14.70 & Sch\\
55814.349 & 12.01 & 13.03 & 13.98 & 15.30 & Sch & 56167.300 & 11.59 & 12.50 & 13.31 & 14.60 & Sch\\
55815.276 & 11.88 & 12.88 & 13.81 & 15.23 & 1.3-m & 56168.310 & 11.54 & 12.42 & 13.26 & 14.55 & Sch\\
55815.316 & -     & 12.87 & -     & -     & Sch & 56169.287 & 11.58 & 12.45 & 13.30 & 14.62 & Sch\\
55816.326 & 11.86 & 12.86 & 13.80 & 15.18 & Sch & 56173.360 & 11.48 & 12.32 & 13.11 & 14.40 & 1.3-m\\
55816.433 & 11.86 & 12.87 & 13.70 & 15.25 & 1.3-m & 56174.338 & 11.41 & 12.24 & 13.03 & 14.30 & 1.3-m\\
55817.244 & 11.87 & 12.90 & 13.85 & 15.24 & Sch & 56178.311 & 11.43 & 12.25 & 13.04 & 14.32 & 1.3-m\\
55818.275 & -     & 12.90 & -     & -     & Sch & 56179.485 & 11.55 & 12.41 & 13.22 & 14.53 & 1.3-m\\
55819.246 & 11.88 & 12.92 & 13.87 & 15.24 & Sch & 56180.346 & 11.55 & 12.41 & 13.24 & 14.54 & 1.3-m\\
55820.276 & -     & 13.16 & -     & -     & Sch & 56181.273 & 11.44 & 12.27 & 13.07 & 14.33 & 60-cm\\
55821.246 & 11.98 & 13.07 & 14.04 & 15.35 & Sch & 56182.268 & 11.43 & 12.26 & 13.06 & 14.33 & 1.3-m\\
55822.238 & 11.88 & 12.93 & 13.90 & 15.30 & Sch & 56183.280 & 11.43 & 12.25 & 13.04 & 11.25 & 60-cm\\
55824.237 & 11.89 & 12.93 & 13.88 & 15.29 & 1.3-m & 56183.393 & 11.44 & 12.27 & 13.07 & 14.33 & 1.3-m\\
55828.281 & 11.75 & 12.77 & 13.71 & 15.13 & Sch & 56192.311 & 11.46 & 12.29 & 13.11 & 14.36 & 60-cm\\
55842.306 & 11.68 & 12.65 & 13.55 & 14.95 & 1.3-m & 56193.308 & 11.54 & 12.39 & 13.21 & 14.51 & 1.3-m\\
55848.297 & 11.76 & 12.79 & 13.69 & 15.09 & 1.3-m & 56193.360 & 11.51 & - & - & - & Sch\\
55864.275 & 12.14 & 13.11 & 14.10 & 15.59 & 2-m & 56194.341 & 11.56 & 12.43 & 13.24 & 14.56 & Sch\\
55865.268 & 11.99 & 12.96 & 13.92 & 15.38 & 2-m & 56195.270 & 11.45 & 12.28 & 13.09 & 14.36 & Sch\\
55866.218 & 12.03 & 12.96 & 13.93 & 15.38 & 2-m & 56208.248 & 11.86 & 12.82 & 13.71 & 15.09 & Sch\\
55890.202 & 12.33 & 13.51 & 14.50 & 15.68 & 60-cm & 56209.251 & 11.98 & 12.97 & 13.87 & 15.26 & Sch\\
55892.232 & 12.39 & 13.43 & 14.50 & 15.98 & 2-m & 56210.242 & 11.98 & 12.95 & 13.86 & 15.22 & Sch\\
55895.212 & 12.64 & 13.80 & 14.75 & 16.07 & Sch & 56212.281 & 11.74 & 12.66 & 13.57 & 14.92 & 60-cm\\
56214.252 & 11.73 & 12.60 & 13.47 & 14.85 & 2-m   & 56513.419 & 11.76 & 12.71 & 13.58 & 14.94 & 60-cm\\
56226.374 & 11.62 & 12.51 & 13.38 & 14.73 & Sch   & 56514.386 & 11.68 & 12.59 & 13.48 & 14.86 & 60-cm\\
56231.280 & 11.79 & 12.76 & 13.71 & 15.07 & 60-cm & 56540.346 & 11.61 & 12.48 & 13.32 & 14.63 & Sch\\
56249.272 & 11.64 & 12.56 & 13.42 & 14.77 & Sch   & 56541.380 & 11.61 & 12.47 & 13.32 & 14.62 & Sch\\
56250.226 & 11.65 & 12.58 & 13.45 & 14.79 & Sch   & 56542.420 & 11.65 & 12.52 & 13.38 & 14.70 & Sch\\
56275.302 & 11.71 & 12.59 & 13.45 & 14.82 & 2-m   & 56543.376 & 11.70 & 12.55 & 13.38 & 14.69 & 2-m\\
56276.259 & 11.64 & 12.51 & 13.35 & 14.67 & 2-m   & 56553.326 & 11.77 & 12.70 & 13.56 & 14.91 & 1.3-m\\
56292.368 & 11.58 & 12.47 & 13.34 & 14.70 & 60-cm & 56577.469 & 11.57 & 12.45 & 13.30 & 14.65 & 60-cm\\
56294.303 & 11.51 & 12.41 & 13.31 & 14.60 & 60-cm & 56578.482 & 11.60 & 12.46 & 13.31 & 14.68 & 60-cm\\
56295.349 & 11.56 & 12.45 & 13.30 & 14.62 & 60-cm & 56604.444 & 11.65 & 12.63 & 13.55 & - & 60-cm\\
56296.327 & 11.61 & 12.49 & 13.37 & 14.72 & 60-cm & 56636.280 & 11.93 & 12.96 & 14.00 & 15.51 & 2-m\\
56309.254 & 11.75 & 12.67 & -     & -     & Sch   & 56655.226 & 12.32 & 13.54 & 14.55 & 15.94 & Sch\\
56312.252 & 11.72 & 12.65 & 13.58 & 15.00 & 2-m   & 56656.234 & 12.34 & 13.56 & 14.57 & 15.91 & Sch\\
56329.210 & 11.71 & 12.67 & 13.57 & 14.93 & Sch   & 56657.212 & 12.39 & 13.59 & 14.60 & 15.97 & Sch\\
56330.218 & 11.82 & 12.81 & 13.75 & 15.13 & Sch   & 56681.239 & 12.25 & 13.43 & 14.44 & 15.87 & Sch\\
56356.261 & 11.52 & 12.41 & 13.30 & 14.58 & 60-cm & 56694.239 & 12.23 & 13.28 & 14.32 & 15.83 & 2-m\\
56369.561 & 11.52 & 12.34 & 13.16 & 14.42 & 2-m   & 56738.547 & 12.38 & 13.55 & 14.51 & 15.81 & Sch\\
56392.487 & 11.50 & 12.35 & 13.17 & 14.46 & Sch   & 56799.494 & 12.12 & 13.25 & 14.16 & 15.58 & Sch\\
56394.432 & 11.48 & 12.35 & 13.17 & 14.42 & Sch   & 56801.344 & 11.93 & 12.92 & 13.88 & 15.32 & 2-m\\
56415.444 & 12.22 & 13.20 & 14.10 & 15.44 & Sch   & 56832.325 & 11.60 & 12.54 & 13.43 & 14.85 & 2-m\\
56417.414 & 11.82 & 12.69 & 13.59 & 14.99 & 2-m   & 56834.319 & 11.67 & 12.59 & 13.53 & 14.94 & 2-m\\
56443.440 & 12.00 & 12.94 & 13.84 & 15.19 & Sch   & 56835.481 & 11.66 & 12.60 & 13.55 & 14.92 & 2-m\\
56444.410 & 11.95 & 12.90 & 13.78 & 15.14 & Sch   & 56837.392 & 11.79 & 12.82 & 13.77 & 15.13 & Sch\\
56478.411 & 11.68 & 12.58 & 13.44 & 14.83 & 2-m   & 56838.374 & 11.82 & 12.85 & 13.79 & 15.14 & Sch\\
56506.411 & 11.72 & 12.58 & 13.40 & 14.73 & 2-m   & 56859.459 & 11.67 & 12.71 & 13.62 & 14.96 & 60-cm\\
56507.408 & 11.85 & 12.71 & 13.55 & 14.88 & 2-m   & 56860.469 & 11.69 & 12.71 & 13.60 & 14.95 & 60-cm\\
56508.446 & 11.90 & 12.77 & 13.60 & 14.93 & 2-m   & 56863.425 & 11.83 & 12.84 & 13.84 & 15.22 & Sch\\
56509.344 & 12.01 & 12.93 & 13.78 & 15.08 & Sch   & 56873.349 & 11.72 & 12.74 & 13.68 & 15.05 & Sch\\
56510.411 & 12.27 & 13.25 & 14.14 & 15.43 & 60-cm & 56874.377 & 11.75 & 12.75 & 13.68 & 15.08 & Sch\\
56511.435 & 12.11 & 13.08 & 13.95 & 15.24 & Sch   & 56888.403 & 11.65 & 12.61 & 13.49 & 14.80 & Sch\\
56511.452 & 12.10 & 13.06 & 13.93 & 15.28 & 60-cm & 56889.338 & 11.64 & 12.61 & 13.47 & 14.81 & Sch\\
56512.441 & 11.90 & 12.85 & 13.70 & 15.03 & 60-cm & 56899.325 & 11.64 & 12.63 & 13.50 & 14.84 & 1.3-m\\
\end{longtable}
\twocolumn

\begin{table}
\caption{Data from $U$ band observations of GM Cep during the period July 2008 - February 2014}
\begin{center}
\begin{tabular}{@{}lllll@{}}
\hline
\hline%
Date & J.D. (24...) & U & Tel & CCD\\ \hline
08 Jul 2008 & 54656.464 & 14.79 & 1.3-m & ANDOR\\
13 Jul 2008 & 54661.428 & 15.61 & 1.3-m & ANDOR\\
24 Jul 2008 & 54672.335 & 15.37 & 1.3-m & ANDOR\\
25 Jul 2008 & 54673.355 & 15.21 & 1.3-m & ANDOR\\
11 Jun 2009 & 54994.581 & 16.08 & 1.3-m & ANDOR\\
14 Jun 2009 & 54997.531 & 16.35 & 1.3-m & ANDOR\\
17 Jun 2009 & 55000.584 & 16.45 & 1.3-m & ANDOR\\
23 Jun 2009 & 55006.517 & 15.83 & 1.3-m & ANDOR\\
01 Jul 2009 & 55014.517 & 16.38 & 1.3-m & ANDOR\\
03 Jul 2009 & 55016.576 & 16.16 & 1.3-m & ANDOR\\
06 Jul 2009 & 55019.512 & 16.47 & 1.3-m & ANDOR\\
09 Jul 2009 & 55022.521 & 16.64 & 1.3-m & ANDOR\\
18 Jul 2009 & 55031.504 & 16.53 & 1.3-m & ANDOR\\
24 Jul 2009 & 55037.526 & 16.59 & 1.3-m & ANDOR\\
31 Jul 2009 & 55044.366 & 16.89 & 1.3-m & ANDOR\\
25 Nov 2009 & 55161.217 & 15.09 & 2-m & VA\\
13 Jul 2010 & 55391.338 & 16.38 & 2-m & VA\\
17 Jul 2010 & 55395.341 & 16.42 & 2-m & VA\\
11 Aug 2010 & 55420.051 & 16.20 & 1.3-m & ANDOR\\
12 Aug 2010 & 55421.367 & 16.13 & 1.3-m & ANDOR\\
25 Aug 2010 & 55434.316 & 16.09 & 1.3-m & ANDOR\\
26 Aug 2010 & 55435.352 & 16.09 & 1.3-m & ANDOR\\
29 Oct 2010 & 55499.295 & 15.68 & 2-m & VA\\
30 Oct 2010 & 55500.238 & 15.43 & 2-m & VA\\
01 Nov 2010 & 55502.259 & 15.50 & 2-m & VA\\
10 Sep 2011 & 55815.276 & 15.99 & 1.3-m & ANDOR\\
31 Oct 2011 & 55866.218 & 16.02 & 2-m & VA\\
03 Sep 2012 & 56174.338 & 14.89 & 1.3-m & ANDOR\\
09 Sep 2012 & 56180.346 & 15.13 & 1.3-m & ANDOR\\
11 Sep 2012 & 56182.268 & 14.92 & 1.3-m & ANDOR\\
22 Sep 2012 & 56193.308 & 15.06 & 1.3-m & ANDOR\\
03 Aug 2013 & 56508.446 & 15.55 & 2-m & VA\\
07 Sep 2013 & 56543.376 & 15.15 & 2-m & VA\\
05 Feb 2014 & 56694.239 & 16.93 & 2-m & VA\\
\hline\hline
\end{tabular}
\end{center}
\label{tab3}
\end{table}

\section{Results and Discussion}

\subsection{Photometric monitoring} 

The results of our photometric CCD observations of GM Cep are summarized in Table 1.  
The columns provide the Julian date (J.D.) of observation, $IRVB$ magnitudes, and the telescope used.
In the column Tel abbreviation 2-m denote the 2-m Ritchey-Chr\'{e}tien-Coud\'{e}, Sch - the 50/70-cm Schmidt, 60-cm - the 60-cm Cassegrain and 1.3-m the 1.3-m Ritchey-Cr\'{e}tien telescope. 
The typical instrumental errors from $IRVB$ photometry are reported in our previous study \cite{sem12}.
In addition, we present in Table 2 data from observations in $U$ filter for the whole period of our photometric monitoring (2008 - 2014).
The values of instrumental errors of $U$ band photometry are in the range 0.04-0.08 mag.

The $UBVRI$ lights curves of GM Cep from all our observations (Semkov \& Peneva \shortcite{sem12} and the present paper) are shown in Fig. 1. 
On the figure triangles denote $I$-band data; squares - $R$-band, circles - $V$-band; diamonds - $B$-band, and the pluses - $U$-band.

\begin{figure*}
\begin{center}
\includegraphics[width=\textwidth]{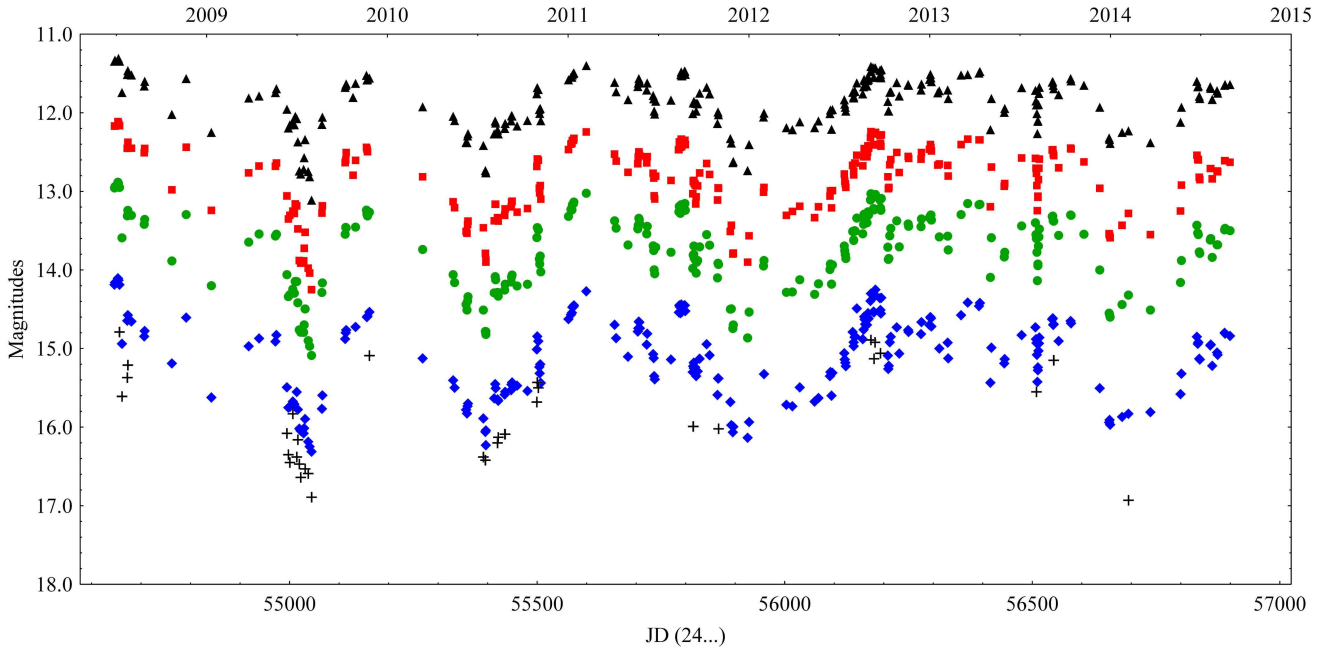}
\caption{$UBVRI$ light curves of GM Cep for the whole period of our photometric monitoring (2008 - 2014)}
 \label{Fig1}
\end{center}
\end{figure*}

The new photometric data showed continued strong brightness variability of GM Cep as the registered in the previous studies \cite{sic08,xiao,sem12,chen}.
Out of deep minimums GM Cep shows significant brightness variations in the time scale of days and months.
In our first paper \cite{sem12} we presented data about two observed deep minimums in brightness.
During the period April 2011 - August 2014 three new well defined minimums in brightness are observed.
The third registered minimum is very extended covering the period from the end of 2011 to mid-2012. 
The fourth minimum has a duration of only 8-9 days and it is registered in August 2013. 
A drop in brightness with 0.74 mag. ($V$) for a period of four days and a rise to the maximum level for the same time was observed.
The fifth minimum is registered in the period from December 2013 to June 2014 and it resembles in duration and amplitude the minimum of 2011/2012. 

\begin{figure}
\begin{center}
\includegraphics[width=\columnwidth]{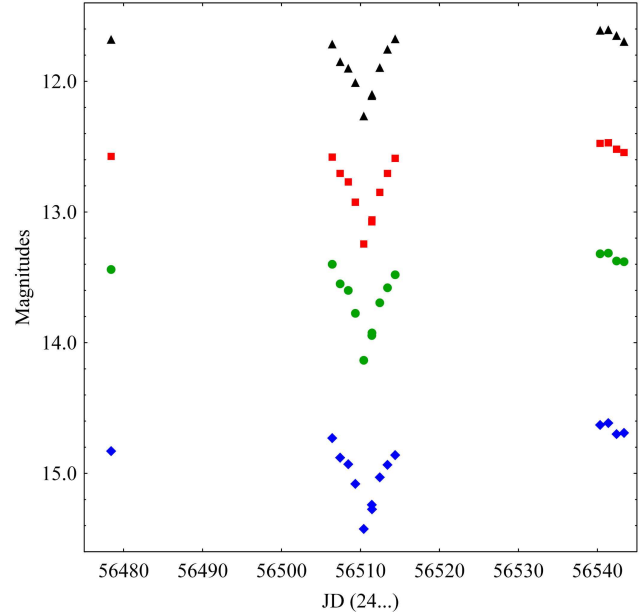}
\caption{$BVRI$ light curves during the deep minimum on August 2013}
 \label{Fig1}
\end{center}
\end{figure}

The summarized results of over six years period of observations show very strong photometric variability.
We have registered five deep minimum in brightness in the light curve of GM Cep.
The first two minimums observed in 2009 and 2010 have a duration of between one and two months, the third (2011/2012) and the fifth (2013/2014) minimum have duration at about half an year, and the fourth minimum (August 2013) has at one week duration (Fig. 2).
Other drops in brightness with duration of about a week have not been surely registered in our photometric study, but the occurrence of such short events cannot be ruled out.
Our photometric data do not confirm the existence of a long-term periodicity, as suggested by Chen \& Hu \shortcite{chenhu}.
Eclipses in the light curve of the star are probably caused by objects of different sizes and densities.
Such objects could be massive dust clumps orbiting the star, inhomogeneous structures of the circumstellar disk or planetesimals at different stages of formation.

Another important result of our study is the change in color of GM Cep at the deep minimums.
Using data from our $UBVRI$ photometry the four color-magnitude diagrams ($U-B/B$, $B-V/V$, $V-R/V$ and $V-I/V$) of the star are constructed and displayed on Fig. 3.
The existence of a turning point of each of the diagrams is seen on the figure.
In accordance with the model of dust-clump obscuration, the observed color reversal is caused by the scattered light from small dust grains.
Generally the star becomes redder when its light is covered by dust clumps on the line of sight.
But when the obscuration rises enough, the part of the scattered light in the total observed light becomes significant and the star color gets bluer.  
For each color, such a turning point occurs at different stellar brightness, for example on $V/B-V$ diagram the turning point occurs at V$\sim$14.0 mag., while on $V/V-I$ diagram at V$\sim$14.6 mag.
As we mentioned in our first paper \cite{sem12} ``the observed change of color indices suggest for existence of a color reversal in the minimum light, a typical feature of the PMS stars from UXor type".
The new data confirm the presence of ``blueing effect" at minimum light and they are independent evidence that the variability of GM Cep is dominated by variable extinction from the circumstellar environment.

\begin{figure*}
\begin{center}
\includegraphics{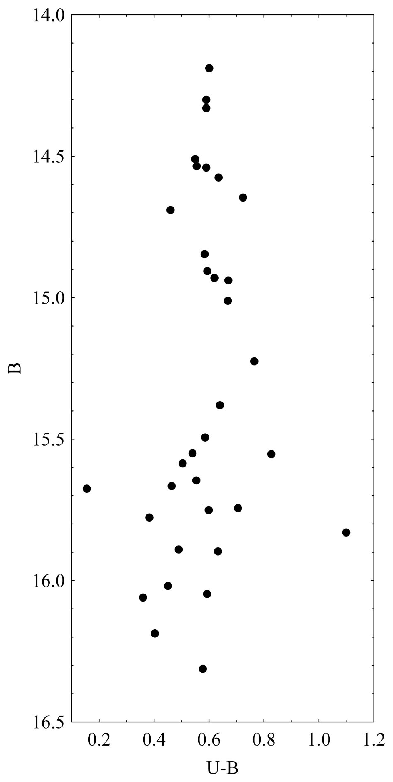}
\includegraphics{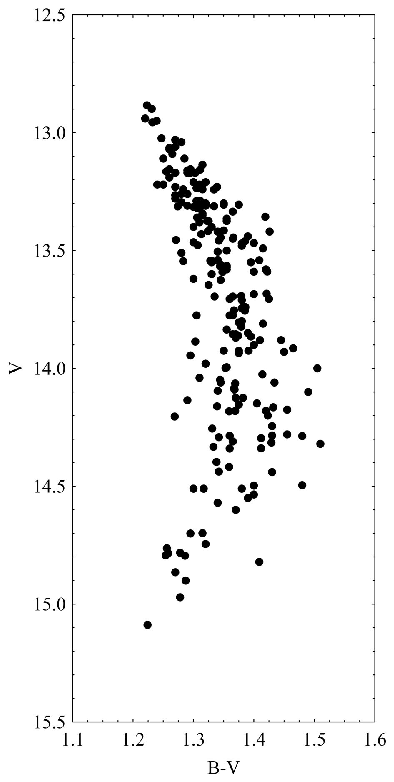}
\includegraphics{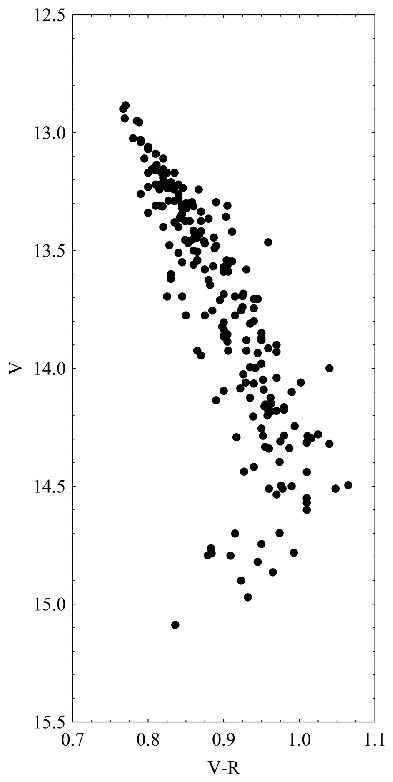}
\includegraphics{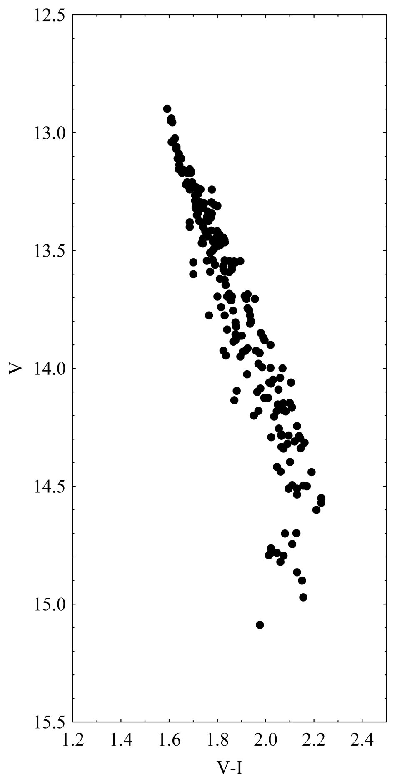}
\caption{The color-magnitude diagrams of GM Cep in the period of observations June 2008 - June 2014}
 \label{Fig2}
\end{center}
\end{figure*}

After analysis of data collected our conclusion is that the photometric properties of GM Cep can be explained by superposition of both: (1) highly variable accretion from the circumstellar disk onto the stellar suffice, and (2) occultation from circumstellar clumps of dust, planetesimals or from features of the circumstellar disk.  
Our photometric results for the period June 2008 - August 2014 suggest that the variable extinction dominates the variability of GM Cep.
In low accretion rates both types of variability can act independently during different time periods and the result is the complicated light curve of GM Cep.

Due to the complex circumstellar environment around PMS stars, such a mixture of different types of photometric variability can be expected.
In recent studies, a similar superposition of the both types of variability is seen on the long-term light curve of others PMS stars: V1184 Tau \cite{sem08,bar}, V1647 Ori \cite{asp09}, V582 Aur \cite{sem13} and V2492 Cyg \cite{hil13}. 
Recently, the results of two long-term photometric studies in the field of NGC 7000/IC 5070 \cite{fin13,pol14} has shown that the eclipsing phenomena are widespread type of variability in among the PMS stars.
It seems that the time variable extinction is characteristic not only of HAEBE and early type CTT stars but is also a common phenomenon during the evolution of all types of PMS stars.

\begin{figure*}
\begin{center}
\includegraphics[width=\textwidth]{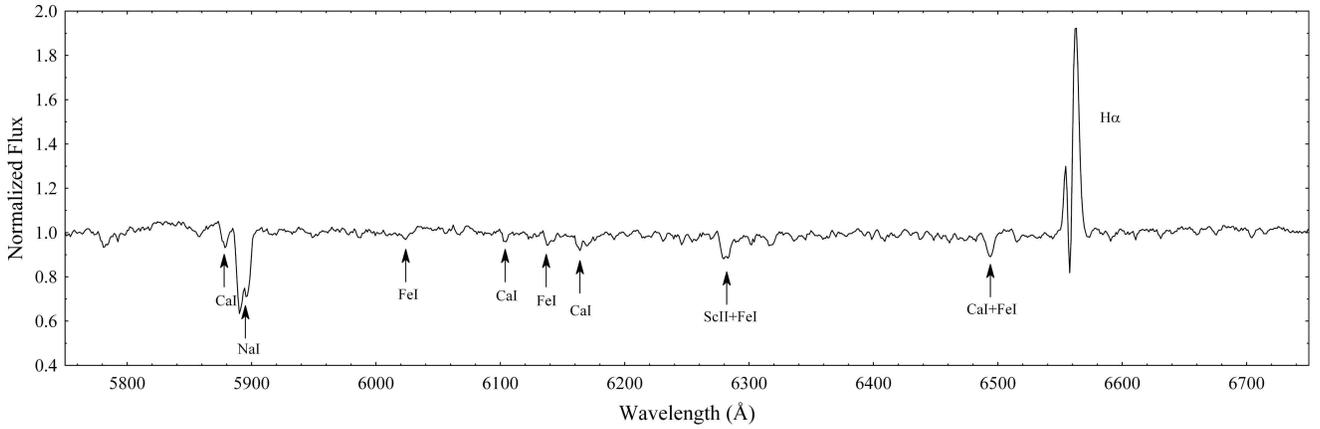}
\caption{Spectrum of GM Cep obtained on 2008 June 27 with the 1.3-m RC telescope in Skinakas Observatory}
 \label{Fig3}
\end{center}
\end{figure*}

\subsection{Spectral data}
The medium-resolution spectrum of GM Cep obtained in Skinakas Observatory is shown in Fig. 4.
At the time of spectral observations (June 2008) the star was at the maximal level of brightness (V $\sim$ 12.9 mag).
The analysis of spectrum was made using the standard procedures in IRAF.
We fits the line profiles with Gaussian and estimate the equivalent width of the lines. 
The spectrum shows the typical of CTT stars absorption lines of iron, calcium, sodium and other metals and a very broad H$\alpha$ emission line.

The double-line profile of the  H$\alpha$ line suggest that the line is formed in a disk-like region \cite{horn}.
There are similarities between the profiles of the H$\alpha$ lines of GM~Cep and some Be/X-ray binary stars, 
e. g. LS~I~+61~303 \cite{zam10}.
The circumstellar disks in Be/X-ray binaries are formed from the fast rotation of the Be star, non-radial pulsations and slow and dense equatorial wind. 
The PMS stars are characterized with strong stellar winds.
In case of GM~Cep, the wind probably form disk-like structure near the surface of the star.
The depth of the central absorption of H$\alpha$ line suggest that the inclination of the star to the line of sight is $i \sim$ 75$^o$ \cite{hanu}.
In Table~\ref{ha} are given the measured parameters of the H$\alpha$ line. 
\begin{table}
\caption{The parameters of the two peaks and the central dip of the H$\alpha$ line. Given are as follows: equivalent width (EW) of the line, full width at half maximum (FWHM) and the radial velocity (V$_{rad}$). } 
\begin{center}
\begin{tabular}{ccccc}
\hline \hline
         & EW	         & FWHM            & V$_{rad}$  &  \\ 
         & [\AA]        & [\AA]          & [km~s$^{-1}$]             & \\
\hline
Blue peak   & -1.09$\pm$0.02       &    3.00$\pm$0.02               &   -392.3$\pm$0.1     &\\
Central dip & +0.23$\pm$0.02       &    1.06$\pm$0.02               &   -231.5$\pm$0.1     &\\
Red peak    & -5.39$\pm$0.02       &    5.05$\pm$0.02               &    5.9$\pm$0.1      & \\
\hline \hline
\end{tabular}\label{tab2}
\end{center}
\label{ha}
\end{table}

For rotationally dominated profiles, the peak separation can be regarded as a measure of 
the outer radius of the H$\alpha$ emitting disk \cite{hua72}: 
 \begin{equation}
      R_{disk} = \frac {G M_* sin^2~i}{(0.5~\Delta V)^2},
  \label{Huang}
  \end{equation}

From the spectrum we estimate $\Delta$ V = 379.4$\pm$0.3 km~s$^{-1}$ (the distance between the blue and red peaks of H$\alpha$). 
This velocity is connected with the outer edge of the disk.
Using mass of the star M$_*$ = 2.1M$_{\odot}$ and inclination angel $i$=75$^o$, we calculate the outer radius
of the H$\alpha$ emitting region to be 10.4$\pm$0.5 R$_{\odot}$.

Using the correlation between the H$\alpha$ velocity wings at 10\% of the maximum (V$_{H\alpha10\%}$) and \.{M}$_{ac}$, we can estimate the accretion rate \cite{nata04}:

\begin{equation}
\log \dot{M}_{ac} = -12.89 (\pm 0.3) + 9.7 (\pm 0.7)\times 10^{-3} V_{H\alpha10\%}
\end{equation}

where V$_{H\alpha10\%}$  is in km~s$^{-1}$ and \.{M}$_{ac}$ is in M$_{\odot}$~yr$^{-1}$.

For measured velocity 633~km~s$^{-1}$ on 10\% of the maximum, the accretion rate is 1.8$\times$10$^{â-7}$ ~M$_{\odot}$~yr$^{-1}$, which is close to the value 3$\times$10$^{â-7}$ ~M$_{\odot}$~yr$^{-1}$, obtained by Sicilia-Aguilar et al. \shortcite{sic08}.

\section{Conclusion}
Photometric and spectral data presented in this paper show the usefulness of systematically monitoring of PMS stars with large amplitude variability. 
On the basis of our photometric monitoring over the past six years, we have confirmed that the variability of GM Cep is dominated by fading events  rather than by bursting events.
The effect of color reversal at the minimum light is evidence of variable extinction from the circumstellar environment.
We plan to continue our photometric monitoring of the star during the next years and strongly encourage similar follow-up observations.

\begin{acknowledgements}
This study was partly supported by ESF and Bulgarian Ministry of Education and Science under the contract BG051PO001-3.3.06-0047. 
The authors thank the Director of Skinakas Observatory Prof. I. Papamastorakis and Prof. I. Papadakis for the award of telescope time. 
The research has made use of the NASA Astrophysics Data System Abstract Service.
\end{acknowledgements}


\begin{thebibliography}{}

  \bibitem[\protect\citename{Aspin et al. }2009]{asp09} Aspin, C., Reipurth, B., Beck, T. L. et al. 2009, ApJ, 692, L67
  
  \bibitem[\protect\citename{Barsunova et al. }2006]{bar} Barsunova, O. Yu., Grinin, V. P., \& Sergeev, S. G. 2006, Astr. Let., 32, 924 
 
  \bibitem[\protect\citename{Bertout }1989]{ber} Bertout, C. 1989, ARA\&A, 27, 351
  
  \bibitem[\protect\citename{Chen et al. }2012]{chen} Chen, W. P., Hu, S. C. -L., Errmann, R. et al. 2012, ApJ, 751, 118
  
  \bibitem[\protect\citename{Chen \& Hu }2014]{chenhu} Chen, W. P. \& Hu, S. C.-L. 2014, IAUS, 293, 74
  
  \bibitem[\protect\citename{Contreras et al. }2002]{con02} Contreras, M. E., Sicilia-Aguilar, A., Muzerolle, J. et al. 2002, AJ, 124, 1585
  
  \bibitem[\protect\citename{Dullemond et al. }2003]{dul03} Dullemond, C. P., van den Ancker, M. E., Acke, B., \&  van Boekel, R. 2003, ApJ, 594, L47
  
  \bibitem[\protect\citename{Findeisen et al. }2013]{fin13} Findeisen K., Hillenbrand L., Ofek E. et al. 2013, ApJ, 768, 93

 	\bibitem[\protect\citename{Grinin et al. }1991]{grin91} Grinin, V. P., Kiselev, N. N., Minikulov, N. Kh., Chernova, G. P., \& Voshchinnikov, N. V. 1991, Ap\&SS, 186, 283
  
  \bibitem[\protect\citename{Hanuschik }1996]{hanu} Hanuschik, R.~W. 1996, A\&A, 308, 170 
  
  \bibitem[\protect\citename{Hartmann \& Kenyon} 1996]{har96} Hartmann, L., \& Kenyon, S. J. 1996, ARA\&A, 34, 207
  
  \bibitem[\protect\citename{Herbst et al. }2007]{her07} Herbst, W., Eisl\"{o}ffel, J., Mundt, R., \& Scholz, A. 2007, in Protostars and Planets V, ed. B. Reipurth, D. Jewitt, \& K. Keil, 297  
 
  \bibitem[\protect\citename{Herbst et al. }1994]{her94} Herbst, W., Herbst, D. K., Grossman, E. J., \& Weinstein, D. 1994, AJ, 108, 1906 
  
  \bibitem[\protect\citename{Hillenbrand et al. }2013]{hil13} Hillenbrand, L. A., Miller, A. A., Covey, K. R. et al. 2013, AJ, 145, 59
  
  \bibitem[\protect\citename{Horne \& Marsh} 1986]{horn} Horne \& Marsh, 1986, MNRAS, 218, 761
  
  \bibitem[\protect\citename{Huang }1972]{hua72} Huang S.-S. 1972, ApJ, 171, 549
  
  \bibitem[\protect\citename{Kun }1986]{kun86} Kun, M. 1986, IBVS, 2961, 1 
 
  \bibitem[\protect\citename{Marschall \& van Altena }1987]{mar87} Marschall, L. A. \& van Altena, W. F. 1987, AJ, 94, 71
    
  \bibitem[\protect\citename{Natta et al. }1997]{nata97} Natta, A., Grinin, V. P., Mannings, V., \& Ungerechts, H. 1997, ApJ, 491, 885 
  
  \bibitem[\protect\citename{Natta et al. }2004]{nata04} Natta, A., Testi, L., Muzerolle, J., Randich, S., Comer\'{o}n, F., Persi, P. 2004, A\&A, 424, 603 
 
  \bibitem[\protect\citename{Poljan\v{c}i\'{c} Beljan et al. }2014]{pol14} Poljan\v{c}i\'{c} Beljan I., Jurdana-\v{S}epi\'{c} R., Semkov E., Ibryamov S., Peneva S., Tsvetkov, M., 2014, A\&A, 568, A49
  
  \bibitem[\protect\citename{Reipurth \& Aspin} 2010]{ra10} Reipurth, B., \& Aspin, C. 2010, in Evolution of Cosmic Objects through their Physical
Activity, eds. H. A. Harutyunian, A. M. Mickaelian, Y. Terzian (Yerevan: Gitutyun), p. 19
             
  \bibitem[\protect\citename{Semkov et al. }2008]{sem08} Semkov, E. H., Tsvetkov, M. K., Borisova, A. P. et al. 2008, A\&A, 483, 537
  
  \bibitem[\protect\citename{Semkov \& Peneva }2012]{sem12} Semkov, E. H. \& Peneva, S. P. 2012, Ap\&SS, 338, 95
  
  \bibitem[\protect\citename{Semkov et al. }2013]{sem13} Semkov, E. H., Peneva, S. P., Munari, U. et al. 2013, A\&A, 556, A60
  
  \bibitem[\protect\citename{Sicilia-Aguilar et al. }2005]{sic05} Sicilia-Aguilar, A., Hartmann, L., Hern\'{a}ndez, J., Brice\~{n}o, C., \& Calvet, N. 2005, AJ, 130, 188
  
  \bibitem[\protect\citename{Sicilia-Aguilar et al. }2008]{sic08} Sicilia-Aguilar, A., Mer\'{i}n, B., Hormuth, F. et al. 2008, ApJ, 673, 382 
 
  \bibitem[\protect\citename{Suyarkova }1975]{su75} Suyarkova, O. 1975, Perem. Zvezdy, 20, 167
 
  \bibitem[\protect\citename{van den Ancker et al. }1998]{van98} van den Ancker, M. E., de Winter, D., \& Tjin A Djie, H. R. E. 1998, A\&A, 330, 145 
   
  \bibitem[\protect\citename{Xiao et al. }2010]{xiao} Xiao, L., Kroll, P., \& Henden, A. 2010, AJ, 139, 1527 	

  \bibitem[\protect\citename{Zamanov et al. }2010]{zam10} Zamanov, R., Stoyanov, K., Mart{\'{\i}}, J., Tomov, N. A., Belcheva, G., Luque-Escamilla, P. L., Latev, G. 2013, A\&A, 559, A87   

\end{thebibliography}

\end{document}